\documentclass{svproc}
\usepackage{url}
\usepackage{amsmath}
\usepackage{graphicx}
\usepackage{subfigure}

\begin{document}
\mainmatter              

\title{Gravitational collapse of Matter in the presence of Scalar field Dark energy}

\titlerunning{Gravitational collapse with Scalar Field Dark Energy}  
%
\author{Priyanka Saha\inst{1} Dipanjan Dey\inst{2}
Kaushik Bhattacharya\inst{1}}
\authorrunning{P Saha, D Dey, K Bhattacharya} 
%
\tocauthor{Priyanka Saha, Dipanjan Dey, and Kaushik Bhattacharya,}
\institute{Department of Physics, Indian Institute of Technology, Kanpur,
Kanpur-208016, India,\\
\email{priyankas21@iitk.ac.in}, \email{kaushikb@iitk.ac.in}
\and
Department of Mathematics and Statistics, Dalhousie University, Halifax, Nova Scotia, Canada B3H 3J5,\\
\email{deydipanjan7@gmail.com}.}

\maketitle              

\begin{abstract}
This study examines the gravitational collapse of an overdense dark matter region in a coupled scalar field dark energy scenario within a flat FLRW background. It finds that, depending on the initial conditions, some overdense regions avoid collapse and expand eternally with the background. The interior overdense region follows a closed FLRW metric, while its boundary is described by generalized Vaidya spacetime, which allows flux across the boundary while preserving the homogeneity of dark energy inside. Dark matter evolves as cold dark matter, but in non-minimal coupling, the modified Klein-Gordon equation alters dark energy evolution. The results highlight the impact of coupled dark energy on dark matter virialization and cosmic structure formation.\dots
\end{abstract}
\section{Introduction}
In the universe, which is largely considered homogeneous and isotropic, cosmic structure formation begins with linear perturbations growing through gravitational instability. Dark matter plays a key role in this process, as overdense regions collapse under gravity. The top-hat collapse model \cite{Baumann:2022mni} describes this evolution using a closed FLRW metric, assuming a matter-only universe, where Newtonian virialization explains equilibrium structures like galaxies by balancing kinetic and potential energy. To investigate the role of dark energy in structure formation, we consider quintessence and phantom-like scalar field dark energy models with minimal and non-minimal matter-dark energy interactions and examine their effects on system evolution. Throughout this paper, we adopt a system of units in which the speed of light and the universal gravitational constant (multiplied by $8\pi$), are both set to unity.
\section{Coupled system}
The action governing the matter-dark energy interacting system is given by, 
\begin{equation}
\mathcal{S} = \int d^4x \, \sqrt{-g} \left[\frac{R}{2} - \rho_m(n,s) - \frac{1}{2} \epsilon \, \partial_{\mu} \phi \, \partial^{\mu} \phi - V(\phi) - \rho_{\text{int}}(n,s,\phi) \right] \, ,
\end{equation}  
where $g$ is the determinant of the metric tensor $g_{\mu\nu}$, and $R$ is the Ricci scalar. Following Brown’s framework \cite{J D Brown_1993}, the energy density of dark matter, $\rho_m(n,s)$, is considered a function of the particle number density $n$ and the entropy density per particle $s$. The $\epsilon=1,-1$ represents the quintessence and phantom-like scalar field, respectively whereas the function $V(\phi)$ represents the potential for scalar field dark energy. Lastly $\rho_{\text{int}}(n,s,\phi)$ is an arbitrary function of $n$, $s$ and $\phi$ for the interaction between dark matter and dark energy, which is absent in case of minimal coupling. 
Now the Einstein equation yields the Einstein tensor $G_{\mu\nu}$ as,
\begin{equation}\label{11}
G_{\mu\nu}=g_{\mu\nu}\left[p_{m}+p_{\text{int}}-\frac{1}{2}\epsilon\partial_{\mu}\phi \partial^{\mu}\phi-V(\phi)\right]+\epsilon\partial_{\mu}\phi \partial_{\nu}\phi+(\rho_{m}+p_{m}+\rho_{\text{int}}+p_{\text{int}})U_{\mu}U_{\nu}\,\, .
\end{equation}
where \begin{eqnarray}
    p_{m}=n\frac{\partial{\rho_{m}}}{\partial{n}}-\rho_{m},~\text{and} ~p_{\text{int}}=n\frac{\partial{\rho_{\text{int}}(n,s,\phi)}}{\partial n}-\rho_{\text{int}}(n,s,\phi).
\end{eqnarray}
\subsection{Spacetime configuration} 
To model the dynamics of the over-dense dark matter region in the presence of coupled dark energy, we use a closed FLRW spacetime embedded in a flat FLRW background:  
\begin{equation} \label{FLRW_combined}
ds^{2}=-dt^{2}+\frac{a^{2}(t)}{1-kr^{2}}dr^{2}+r^{2}a^{2}(t)(d\theta^{2}+\sin^{2}\theta d\Phi^{2})\,,
\end{equation}
where a close FLRW metric with $k=1$ ensures a turnaround scenario, while the background scale factor follows a flat FLRW evolution with $k=0$. The boundary of the over-dense region is modeled by a generalized Vaidya spacetime, which allows the external flux.
\begin{eqnarray}
    dS^2= -\left(1-\frac{2M(r_v , v)}{r_v}\right)dv^2 - 2dv dr_v + r_v^2 d\Omega^2 .
\end{eqnarray}
Matching the internal closed FLRW spacetime smoothly with the external generalized Vaidya spacetime, we obtain:
\begin{eqnarray}
\label{Massderivative}
M(r_v,v)_{,r_v}&=&\frac{F}{2r_{v}}+\frac{r_{v}^2\ddot{a}}{a}\,\, .
\end{eqnarray}
Here, $F$ denotes the Misner-Sharp mass of the internal collapsing spacetime, which must satisfy the following condition at the boundary:
\begin{equation}
F(t,r)=2M(r_v,v)\,\,.
\label{FM1}
\end{equation}
The matter flux at the boundary depends on the scale factor and the Misner-Sharp mass $F$, which is only time-dependent due to the spatial homogeneity of the internal spacetime. Any non-zero internal pressure, $p = - \frac{\dot{F}}{\dot{R}R^2}$, ensures a nonzero matter flux at the boundary, with inward flux during expansion, and outward flux during collapse, which maintains the homogeneity of the dark-energy. This can be also seen from plotting the mass profile of the system \cite{Saha:2023zos}. Using the FLRW metric the cosmological equations become: 
\begin{eqnarray}\label{one}
\frac{3\dot{a}^2}{a^2}+\frac{3k}{a^2}=\left(\rho_{m}+\frac{1}{2}\epsilon\dot\phi^2+V(\phi)+\rho_{\text{int}}\right)\, ,
\end{eqnarray}
\begin{eqnarray}\label{two}
\frac{2\ddot{a}}{a}+\frac{\dot{a}^2}{a^2}+\frac{k}{a^2}=-\left(p_{m}+\frac{1}{2}\epsilon\dot\phi^2-V(\phi)+p_{\text{int}}\right)\, .
\end{eqnarray}
\begin{eqnarray}\label{Klein1}
\epsilon\ddot{\phi}+3H\epsilon\dot{\phi}+\frac{\partial V}{\partial\phi}+\frac{\partial \rho_{\text{int}}}{\partial\phi}=0,
\end{eqnarray}
\begin{eqnarray}\label{mEq}
\dot{\rho}_{m} + 3\frac{\dot{a}}{a}(\rho_{m}+p_{m})=0.
\end{eqnarray}
\subsection{Construction of the Model}
We require explicit functional forms of the interaction part to solve the system. We begin by describing the system using an autonomous set of differential equations \cite{Boehmer:2015kta} involving three dynamic variables:
\begin{equation}
\dot{x} = f_{1}(x, y, z), \quad \dot{y} = f_{2}(x, y, z), \quad \dot{z} = f_{3}(x, y, z),
\end{equation}
where the functions $f_i$ depend only on $x$, $y$, and $z$.
Now the background evolution in a flat FLRW spacetime is parameterized using the phase-space variables:
\begin{equation}
X = \frac{\dot{\phi}}{\sqrt{6}H}, \quad Y = \frac{\sqrt{V(\phi)}}{\sqrt{3}H}, \quad Z = \frac{\rho_{\text{int}}}{3H^{2}}, \quad \sigma = \frac{\sqrt{\rho_{m}}}{\sqrt{3}H},
\end{equation}
which satisfies the Friedmann constraint:
\begin{equation}
1 = \sigma^{2} + X^{2} + Y^{2} + Z.
\end{equation}
This allows $\sigma$ to be eliminated in favor of other variables. The field equations can be rewritten in terms of $X$, $Y$, $Z$, and their derivatives with respect to $Hdt$ as well as $A$, and $B$. Here
\begin{equation}
A = \frac{p_{\text{int}}}{2H^{2}}, \quad B = \frac{1}{\sqrt{6}H^{2}}\frac{\partial\rho_{\text{int}}}{\partial\phi}.
\end{equation}
To ensure closure of the system, we define $\rho_{\text{int}}$ and $p_{\text{int}}$ as
\begin{equation}
\rho_{\text{int}} = \gamma\rho^{\alpha}_{m}e^{-\beta\phi}, \quad p_{\text{int}} = (\alpha-1)\rho_{\text{int}} \, ,
\end{equation}
where $\gamma$, $\alpha$, and $\beta$ are real constants. We consider $V(\phi) = V_{0}e^{-\lambda\phi}$ for both quintessence-like and phantom-like fields. For dust-like dark matter with Eq. (\ref{mEq}) and $p_{m}=0$, its energy density follows, $\rho_{m}=\rho_{m_{0}}/a^{3}$, where $\rho_{m_{0}}$ is the initial energy density. Now, using these functions the system can be solved. 
\begin{figure*}
    \centering
    \begin{minipage}{0.49\textwidth}
        \centering
        \includegraphics[height=30mm,width=\linewidth]{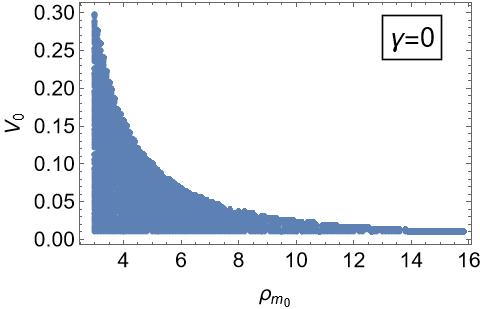}
        \includegraphics[height=30mm,width=\linewidth]{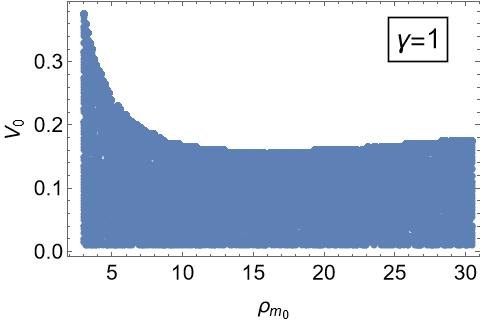}
    \end{minipage}
    \begin{minipage}{0.49\textwidth}
        \centering
        \includegraphics[height=30mm,width=\linewidth]{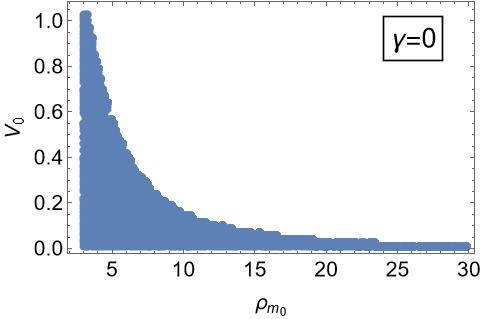}
        \includegraphics[height=30mm,width=\linewidth]{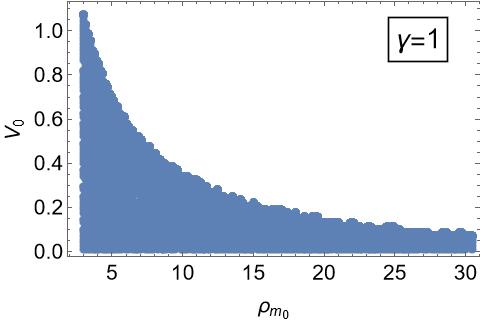}
    \end{minipage}

    \caption{Region plots of $V_{0}$ vs. $\rho_{m_{0}}$ for different $\gamma$ values, with other parameters fixed ($\alpha=\beta=1$). The left two figures correspond to the phantom field and the right two to the quintessence field. Here, $\gamma = 0$ represents minimal coupling, and $\gamma = 1$ indicates non-minimal coupling.}
    \label{Regionplotforvariationgamma}
\end{figure*}

\section{Results}
We identified a bounded \(\rho_{m_0}\)-\(V_0\) region for both minimal and non-minimal coupling scenarios, where only values within this region lead to collapse; otherwise, the system expands eternally. A higher scalar field potential requires a lower initial dark matter density for collapse, showing an inverse relationship. The bounded region is different for quintessence and phantom field as shown in Fig. (\ref{Regionplotforvariationgamma}), and also differs in minimal and non-minimal coupling cases. Non-minimal coupling enlarges collapse-permitting parameter space while restricting eternal expansion. Similar effects arise from variations in \(\beta\) and \(\alpha\) highlighting the impact of dark energy-matter interaction \cite{Saha:2024xbg} on structure formation.

%
%

\end{document}